# Hydrogen isotopic evidence for early oxidation of silicate Earth

Kaveh Pahlevan[1*], Laura Schaefer[1], Marc M. Hirschmann[2]

1. School of Earth and Space Exploration, Arizona State University, Tempe, AZ, 85287, USA

2. Department of Earth Sciences, University of Minnesota, Minneapolis, MN, 55455, USA

*To whom correspondence should be addressed:

Email: kaveh.pahlevan@asu.edu

Tel: +1 (480) 401 8584

Fax: +1 (480) 965 8102

6 Figures

**Abstract**

The Moon-forming giant impact extensively melts and partially vaporizes the silicate Earth and delivers a substantial mass of metal to Earth's core. The subsequent evolution of the magma ocean and overlying atmosphere has been described by theoretical models but observable constraints on this epoch have proved elusive. Here, we report thermodynamic and climate calculations of the primordial atmosphere during the magma ocean and water ocean epochs respectively and forge new links with observations to gain insight into the behavior of volatiles on the Hadean Earth. As accretion wanes, Earth's magma ocean crystallizes, outgassing the bulk of its volatiles into the primordial atmosphere. The redox state of the magma ocean controls both the chemical composition of the outgassed volatiles and the hydrogen isotopic composition of water oceans that remain after hydrogen escape from the primordial atmosphere. The climate modeling indicates that multi-bar $H_2$-rich atmospheres generate sufficient greenhouse warming and rapid kinetics resulting in ocean-atmosphere $H_2O$-$H_2$ isotopic equilibration. Whereas water condenses and is mostly retained, molecular hydrogen does not condense and can escape, allowing large quantities (~$10^2$ bars) of hydrogen – if present – to be lost from the Earth in this epoch. Because the escaping inventory of H can be comparable to the hydrogen inventory in primordial water oceans, equilibrium deuterium enrichment can be large with a magnitude that depends on the initial atmospheric $H_2$ inventory. Under equilibrium partitioning, the water molecule concentrates deuterium and, to the extent that hydrogen in other forms (e.g., $H_2$) are significant species in the outgassed atmosphere, pronounced D/H enrichments (~1.5-2x) in the oceans are expected from equilibrium partitioning in this epoch. By contrast, the common view that terrestrial water has a carbonaceous chondritic source requires the oceans to preserve the isotopic composition of that source, undergoing minimal D-enrichment via equilibration with $H_2$ followed by hydrodynamic escape. Such minimal enrichment places upper limits on the amount of primordial atmospheric $H_2$ in contact with Hadean water oceans and implies oxidizing conditions (log$fO_2$>IW+1, $H_2/H_2O$<0.3) for outgassing from the magma ocean. Preservation of an approximate carbonaceous chondrite D/H signature in the oceans thus provides evidence that the observed oxidation of silicate Earth occurred before crystallization of the final magma ocean, yielding a new constraint on the timing of this critical event in Earth history. The seawater-carbonaceous chondrite "match" in D/H (to ~10-20%) further constrains the prior existence of an atmospheric $H_2$ inventory – of any origin – on post-giant-impact Earth to <20 bars, and suggests that the terrestrial mantle supplied the oxidant for the chemical resorption of metals during terrestrial late accretion.

## 1. Introduction

The composition and origin of Earth's early atmosphere has been debated since at least the mid-twentieth century (Brown, 1949). Recent interest arises from a desire to understand climate on the early Earth (Wordsworth and Pierrehumbert, 2013a) as well as the environment that led to abiogenesis (Kasting, 2014), and because the volatile history of Earth gives insight into the origin and evolution of the planet more generally. Here, we develop the connection between the history of the fluid envelope and that of the silicate Earth. We use ideas about Earth accretion and insights that they yield for the origin and history of terrestrial volatiles. The focus is on Earth's primordial atmosphere, a unique reservoir in Earth history that links the energetic process of planetary accretion via giant impact to the emergence of Earth's oceans and Hadean climate. We use "steam atmosphere" to describe any atmosphere prevented from condensation by internal heat in which $H_2O$ is an important component, even those in which other gases (e.g., $H_2$) are more abundant by number.

A question closely related to the atmospheric composition of the early Earth is the oxygen fugacity ($fO_2$) of the magma ocean from which the primordial atmosphere was outgassed (Elkins-Tanton, 2008). Plausible $fO_2$ of magma ocean outgassing range from reducing ($H_2$-CO-rich) to oxidizing ($H_2O$-$CO_2$-rich) atmospheres (Hirschmann, 2012). In analogy with the volatile inventory on modern Earth, prior work has assumed that the primordial atmosphere of early Earth was $H_2O$-$CO_2$-rich (Abe and Matsui, 1988; Kasting, 1988; Lebrun et al., 2013; Salvador et al., 2017). However, given a lack of knowledge about the oxygen fugacity of outgassing, reducing primordial outgassed atmospheres are difficult to rule out (Abe et al., 2000; Hirschmann, 2012). Models and measurements of early Earth oxygen fugacity yield contradictory evidence: whereas the $fO_2$ of metal-silicate equilibration is necessarily reducing ($\log fO_2 < IW-2$) due to the co-existence of metals and implies $H_2$-CO-rich gas mixtures (Gaillard et al., 2015; Wade and Wood, 2005), the oldest terrestrial samples are characterized by much higher $fO_2$ consistent with the modern oxidized mantle and suggest an $H_2O$-$CO_2$-rich atmosphere (Delano, 2001; Nicklas et al., 2018; Trail et al., 2011). Some process apparently oxidized the silicate Earth during or shortly after core formation. Because the nature and timing of this process are as yet unclear, the chemical composition and oxidation state of the early Earth's atmosphere remain unknown.

Here, by linking the hydrogen isotopic composition of the terrestrial oceans to the chemical composition of the primordial atmosphere, we articulate new constraints on the oxygen fugacity of magma ocean outgassing and subsequent processes on the Hadean Earth. Because this model is constrained by isotopic observations, it has the potential to yield new insights into early Earth evolution. Several features of Earth evolution imply that the D/H composition of the oceans reflects early atmospheric processes: (1) most of the water gained by Earth during planet formation was already accreted at the time of the Moon-forming giant impact and therefore participated in the terminal terrestrial magma ocean (Fischer-Gödde and Kleine, 2017; Greenwood et al., 2018), (2) most (>70%) of the water initially dissolved in the magma ocean was outgassed during solidification, rendering the steam atmosphere the dominant exchangeable water reservoir on early Earth (Elkins-Tanton, 2008), (3) the steam atmosphere was too short-lived for significant water loss via hydrodynamic escape before the condensation of the oceans (Hamano et al., 2013; Massol et al., 2016), (4) the residence time of water in the terrestrial oceans with respect to subduction and the deep water cycle is long, of order $\sim 10^{10}$ years (van Keken et al., 2011). Together, these observations suggest that most hydrogen atoms currently residing in the oceans experienced the magma ocean and its aftermath and carry isotopic memory from early epochs. The inference that the modern oceans reflect the isotopic composition of Hadean oceans – and, indeed, the bulk silicate Earth – is supported by measurements on early Archean samples with D/H values for seawater identical to modern values to within a few percent (Pope et al., 2012). Percent-level variations in ocean D/H can arise due to exchange of water with the solid Earth due to plate tectonic processes (Kurokawa et al., 2018; Lécuyer et al., 1998) but here we are interested in large magnitude D/H variations (~1.5-2x) that can arise due to early atmospheric processes (see §4.3).

The isotopic composition of the oceans is determined by deuterium partitioning between $H_2O$ and $H_2$ and the contrasting histories of these molecules in the planetary environment. Hydrogen ($^1H$) and deuterium ($^2D$) in water and molecular hydrogen experience distinct vibrational frequencies due to different bond strengths associated with the O-H and H-H stretch. These distinct bonding environments result in deuterium being concentrated into water molecules with molecular hydrogen deuterium-depleted at equilibrium, especially at low temperatures, i.e., isotopic fractionation occurs between the water ocean and $H_2$-rich atmosphere. Hence, even in the absence of $HD/H_2$ mass fractionation during the escape process (Zahnle et al., 1990), loss of molecular hydrogen from the early Earth can enrich planetary water in deuterium due to equilibrium partitioning because light isotopes are preferentially concentrated into the escaping atmosphere relative to the oceans (Genda and Ikoma, 2008). As on modern Earth, water vapor is expected to be confined to the lower atmosphere via condensation below the tropopause (Wordsworth and Pierrehumbert, 2013b) and retained on the Hadean Earth whereas hydrogen in non-condensable forms (e.g., $H_2$) can traverse the tropopause and undergo large-scale escape (Catling et al., 2001; Hunten and Strobel, 1974; Lammer et al., 2018; Tian et al., 2005). To the extent that the isotopic composition of the escaping gas was distinct from that of the oceans, isotopic evolution would have taken place.

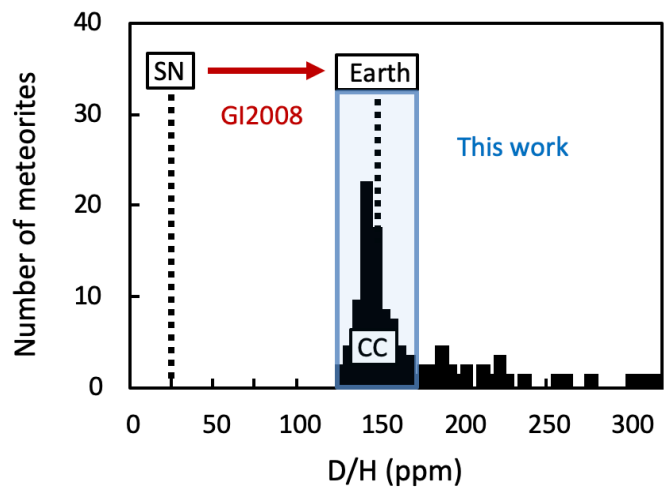

Figure 1 – **Origin of deuterium-to-hydrogen on Earth.** Measured values of D/H in carbonaceous chondrites (CC) (Robert, 2003), the inferred values

of the bulk silicate Earth (149+/-3 ppm) (Lécuyer et al., 1998) and the solar nebula (SN) (26+/-7 ppm) (Mahaffy et al., 1998). A nebular source for terrestrial water requires a ~six-fold D/H enrichment due to $H_2$ escape over several Ga (Genda and Ikoma, 2008), whereas a carbonaceous chondrite source (shaded region) disallows significant (~2x) deuterium enrichment via hydrogen escape if the source signature is to be preserved. Here, we develop the consequences of such preservation for the early Earth.

That the hydrogen isotopic composition of the terrestrial oceans retains memory of early epochs permits its use in constraining early atmospheric processes, pending knowledge of the D/H of the source. Indeed, the source of Earth's oceans are commonly inferred from its D/H composition in comparison with early Solar System reservoirs such as comets, asteroids, and the solar nebula. Based on the hydrogen and nitrogen isotopic evidence, the major terrestrial volatiles (C, N, H) are commonly inferred to be sourced primarily from carbonaceous chondrites (Alexander et al., 2012; Halliday, 2013; Marty, 2012). The close match (to 10-20%, Fig. 1) between terrestrial D/H and the carbonaceous chondrite distribution peak – if not genetic – must be relegated to coincidence with low a priori probability (Lécuyer et al., 1998). Moreover, other potential sources for Earth's major volatiles face severe difficulties. The impact probability of comets with Earth is small (~ppm) such that the inferred mass of cometary material can only contribute <10% of a terrestrial ocean, even with complete volatile retention during impact (Dauphas and Morbidelli, 2014). A nebular origin for terrestrial water has been discussed (Ikoma and Genda, 2006) but the solar neon inventory of the Earth only implies ingassing of a small fraction (<10%) of Earth's water inventory into an early magma ocean in the presence of the solar nebula (Wu et al., 2018). Moreover, a nebular source for terrestrial water would require a ~6x deuterium enrichment in the oceans (Fig. 1) due to gentle $H_2$ escape over several Ga (Genda and Ikoma, 2008), whereas the nearly modern ocean D/H at 3.8 Ga (Pope et al., 2012) points to more limited primordial isotopic fractionation and a minor role for any nebular contribution. Finally, the nitrogen isotopic evidence argues for a carbonaceous chondrite source for Earth's major volatiles (Alexander et al., 2012; Halliday, 2013; Marty, 2012). We adopt the common view that the terrestrial oceans have a primarily carbonaceous chondritic source and show that the preservation of such a chondritic D/H signature in the oceans places an upper limit on $H_2$ inventories on the Hadean Earth, requires oxidizing conditions for magma ocean outgassing ($H_2/H_2O<0.3$), and suggests a limited role for $H_2$ production via the iron-water reaction during late accretion.

The outline of the paper is as follows. In §2, we estimate equilibration timescales between the magma ocean and primordial atmosphere and present calculations that relate the chemical composition of the outgassed atmosphere to the oxidation state of the magma ocean. In §3, we motivate and introduce a climate model to calculate greenhouse warming by the primordial $H_2$ inventory in the subsequent water ocean epoch. In §4, we describe the results of the climate model for the isotopic evolution of the oceans due to equilibration and loss of a primordial $H_2$ inventory. In §5, we discuss the implications of these results for the oxidation state of the magma ocean and the oxidant for late accretion, and in §6, we summarize and conclude. The envisioned sequence explored in this paper is summarized in Figure 2.

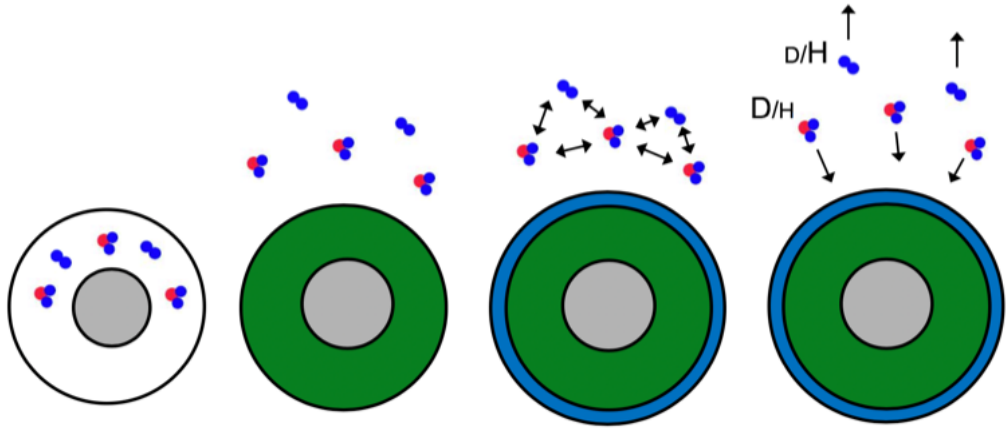

**Figure 2 – Behavior of hydrogen on Earth after the Moon-forming giant impact.** (a) A deep magma ocean dissolves most of the hydrogen accreted to Earth, (b) crystallization of the magma ocean leads to outgassing of most of exchangeable hydrogen with $H_2/H_2O$ determined by oxygen fugacity of last equilibration (§2.2), (c) condensation of the oceans (§3.1) and low-temperature (~300-600K) D/H equilibration (§4.3) leads to deuterium-enrichments ($H_2O$) and depletions ($H_2$) in co-existing species, (d) retention of water via condensation and loss of $H_2$ via hydrodynamic escape produces deuterium enrichment in the oceans whose magnitude depends on the initial $H_2/H_2O$ of the outgassed atmosphere.

## 2. Magma ocean outgassing

Magma oceans arise in the early Solar System through various processes (Elkins-Tanton, 2012). The Moon-forming giant impact extensively melts the silicate Earth and leaves the accreting planet with ~99% of its final mass (Pahlevan and Morbidelli, 2015). During the ensuing magma ocean crystallization period, terrestrial water transitions from being predominantly dissolved in the magma ocean to primarily outgassed into the steam atmosphere, subsequent condensation of which forms the early terrestrial oceans (Elkins-Tanton, 2008; Hamano et al., 2013) (see §3.1). This event is considered the major volatile processing event in Earth history, after which the abundance and initial distribution of water is largely established and planetary processes transition from the accretionary to the geological regime. A growing body of evidence suggests that most water in the silicate Earth was initially concentrated near the surface and that Earth's deep water cycle has been characterized by a net influx of water into the mantle over geologic time, consistent with extensive early outgassing (Kendrick et al., 2017; Korenaga, 2008; Kurokawa et al., 2018; van Keken et al., 2011). In this section, we first justify the use of equilibrium thermodynamics in calculating outgassed atmospheric compositions (§2.1) and then discuss the dependence of the outgassed gaseous composition on the redox state of the magma ocean at the time of last equilibration with the atmosphere (§2.2).

2.1. Magma ocean-atmosphere equilibration timescales

To justify the use of equilibrium thermodynamics, we first examine equilibration times between the magma ocean and the overlying atmosphere. On the modern Earth, the timescale for $pCO_2$ equilibration with the oceans is $\sim 10^2$ years (Archer et al., 2009), but no equivalent empirical estimate exists for magma ocean-atmosphere equilibration. Using a boundary-layer analysis (Hamano et al., 2013), we estimate the timescale for magma ocean-atmosphere equilibration assuming diffusion through the magma surface boundary layer, rather than ascent of bubble plumes, dominates the equilibration process. We consider a schematic sequence in which thermal boundary layers form at the magma ocean surface and are peeled away by negative buoyancy. The equilibration timescale can then be estimated:

$$\tau_{eq} = \tau_{BL} N = (\delta_T^2/\kappa) \times (z/\delta_C) \quad (1)$$

where $\tau_{BL}$ is the timescale for formation of a thermal boundary layer by thermal diffusion and N is the number of formation and buoyant destruction cycles before the entire magma ocean mass is processed through the surface boundary layer, $\delta_T$ and $\delta_C$ are the thermal and chemical boundary layer thicknesses, respectively, $\kappa$ is the thermal diffusivity of the magma (cm$^2$ s$^{-1}$), and z is magma ocean depth. We can relate the thickness of the chemical boundary layer ($\delta_C$) to that of the thermal boundary layer ($\delta_T$) via scaling: $(\delta_C/\delta_T) = (D/\kappa)^{1/2}$ where D is the atomic diffusivity for water in magma (cm$^2$ s$^{-1}$). For parameter choices, we adopt a thermal diffusivity ($\kappa=2\times10^{-3}$ cm$^2$ s$^{-1}$) appropriate to peridotite liquid (Lesher and Spera, 2015) and an atomic diffusivity ($D_w=1.4\times10^{-4}$ cm$^2$ s$^{-1}$) appropriate for a basaltic magma with ~1 wt% water at 2000 K (Zhang et al., 2007). Substitution of these parameters yields an estimate for equilibration timescales:

$$\tau_{eq} = 3 \times 10^3 \, years \left(\frac{\delta_T}{1 cm}\right)\left(\frac{z}{500 \, km}\right) \quad (2)$$

This result suggests that equilibration with the atmosphere by processing the magma ocean through a chemical boundary layer is rapid relative to the crystallization timescale, which is estimated to be ~10$^6$ years (Lebrun et al., 2013). Although more work is required to quantify the competing role of outgassing via bubble plumes in accommodating the supersaturation near the magma surface (Ikoma et al., 2018), this calculation suggests that boundary layer diffusion alone may be sufficiently rapid to motivate the equilibrium assumption. We therefore expect the magma ocean and primordial atmosphere to evolve as a coupled thermochemical system such that the properties of the system at equilibrium (e.g., the oxygen fugacity) reflect the properties of the components (e.g., the atmospheric composition).

2.2. Primordial atmospheric compositions

A fundamental parameter governing equilibrium compositions of outgassed atmospheres is the oxygen fugacity ($fO_2$). As long as the redox buffering capacity of the magma ocean exceeds that of the outgassed atmosphere, the magma ocean determines the $fO_2$ of the atmosphere with which it equilibrates (Hirschmann, 2012). Volatile outgassing continues until the end of magma ocean crystallization, when the formation of a meters-thick solid chill crust isolates the newly formed atmosphere from rapid exchange with the silicate Earth. Therefore, the initial composition of the atmosphere during the Hadean is dictated by last equilibration with the magma ocean.

Despite its importance to planetary evolution, the $fO_2$ at the magma ocean-atmosphere interface is not well-constrained. There are two end-members: (1) in analogy with the modern Earth mantle, the magma ocean-primordial atmosphere system may be chemically oxidized ($\log fO_2 \approx$ QFM), with water vapor and carbon dioxide dominant (Abe and Matsui, 1988; Kasting, 1988; Lebrun et al., 2013; Salvador et al., 2017). However, (2) the terrestrial magma ocean – having held metallic droplets in suspension – may also be much more reducing ($\log fO_2 \approx$ IW-2) at the surface where equilibration with the atmosphere takes place. Under reducing conditions, $H_2$ and CO are the dominant H- and C-bearing gaseous species (Fig. 3). A remarkable feature of magma oceans is that the expected range of possible $fO_2$ values spans the transition from the reducing ($H_2$-CO-rich) to oxidizing ($H_2O$-$CO_2$-rich) gas mixtures, indicating a rich volatile-processing history during early planetary evolution.

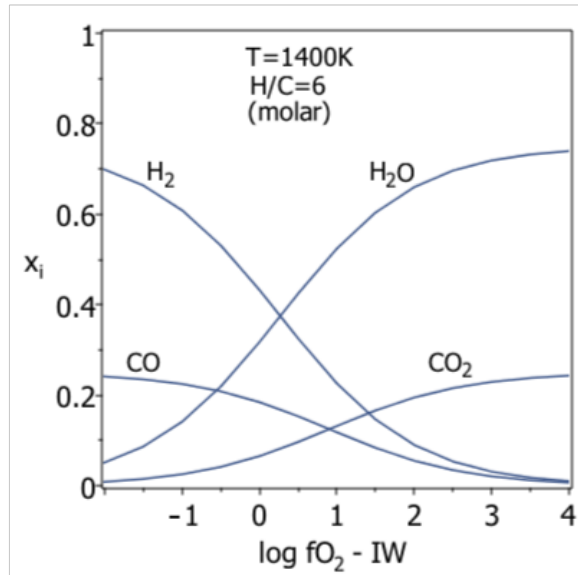

**Figure 3 – High-temperature equilibrium outgassed atmospheres.** The mole fraction of vapor species is calculated as a function of oxygen fugacity ($fO_2$) relative to the iron-wüstite (IW) buffer at an equilibrium temperature of 1,400 K. Parameters for the IW buffer are given in (Frost, 1991). Thermodynamic data for gaseous species ($H_2O$-$H_2$-$CO$-$CO_2$) are adopted from (Chase et al., 1985). The lower end of the range of $\log fO_2$ (IW-2) characterizes oxygen fugacity of a magma ocean in equilibrium with a suspension of metallic droplets, whereas the upper end of the range (IW+4≈QFM) corresponds to the redox state of the modern Earth mantle. Because hydrogen and carbon speciation reactions depend on $fO_2$ alone, different H/C ratios yield similar results for $H_2/H_2O$ and $CO/CO_2$.

Whereas an oxidizing ($H_2O$-$CO_2$-rich) outgassed gaseous envelope experiences minimal chemistry upon cooling and maintains its molecular composition, evolution of a reducing ($H_2O$-$CO$-$H_2$-rich) envelope may involve more significant chemical transformations. As the reducing envelope cools following magma ocean crystallization, the equilibrium reaction ($3H_2+CO \Leftrightarrow CH_4+H_2O$) shifts to the right, potentially converting the outgassed mixture into a methane-rich atmosphere before condensation of the oceans (Schaefer and Fegley, 2010). Such internal transformation changes the molecular composition but not atomic abundances of the envelope. However, we consider $CH_4$ to be – at most – a transient molecule in the primordial atmosphere because it is unstable with respect to photolysis via Lyman $\alpha$ emission. The photolysis of methane under the influence of the UV flux of the young Sun occurs rapidly, at a rate of ~1 bar/Myrs (Kasting, 2014), with the carbon oxidized to $CO/CO_2$ and the hydrogen reverting to molecular form ($H_2$) before escaping, a somewhat more oxidizing analog to the modern atmosphere of Titan. The photochemical stability of $H_2$ suggests that the reducing power inherited from the magma ocean is primarily carried by – and lost via the escape of – molecular hydrogen.

Oxidizing ($H_2O$-$CO_2$-rich) atmospheres and their associated climates have previously been described (Wordsworth and Pierrehumbert, 2013b) but equivalent reducing ($H_2O$-$CO$-$H_2$-rich) atmospheres and climates have not. Our approach in the rest of the paper is to consider climates for a range of atmospheres more reducing than the oxidizing end-member (§3), link the chemical composition of outgassed atmospheres to the ocean D/H, and articulate constraints on reduced gas abundances in Earth's Hadean atmosphere (§4).

**3. Primordial climate**

Calculation of planetary climate after magma ocean crystallization requires specification of the mass and molecular composition of the early atmosphere and presence or absence of water oceans. In this section, we discuss the timescales for formation of water oceans (§3.1) and introduce a model for calculating primordial climates (§3.2).

3.1. Rapid formation of water oceans

Following magma ocean crystallization, the outgassed atmosphere is no longer thermally buffered and its short-term evolution can be considered decoupled from the solid Earth. We consider outgassed water inventories of 1-2 modern oceans equivalent, which is within the range of inferred abundances for the bulk silicate Earth (Hirschmann, 2006) and which can arise from a magma ocean with ~0.1 wt% $H_2O$ (Elkins-Tanton, 2008). Crystallization reduces the geothermal heat flux and leads to an atmospheric heat flux below the runaway greenhouse threshold (Hamano et al., 2013), at which point the stable climate state requires condensation of the steam into oceans. On a rapid timescale relative to escape (>0.1-10 Ma, Fig. 5), the atmosphere behaves as a closed system. Heat inherited from the magma ocean powers atmospheric secular cooling for a timescale:

$$\tau_{cool} = \frac{P_T C_P \Delta T}{g \sigma_{SB} T_E^4} \qquad (3)$$

with $P_T$ the total atmospheric pressure, $C_P$ the thermal heat capacity, $\Delta T$ the atmospheric temperature difference between the magma ocean and water ocean epochs, g the surface gravity, $\sigma_{SB}$ the Stefan-Boltzmann constant, and $T_E$ the effective temperature for cooling. In a steam atmosphere where opacity is dominated by water vapor, the outgoing radiation flux ($\sigma_{SB} T_E^4$) has a lower limit (~300 W/m²) due to condensation (Nakajima et al., 1992). Adopting $P_T$=270-540 bars appropriate for a steam atmosphere with 1-2 modern oceans of water and $\Delta T=10^3$ K for cooling from a steam atmosphere (1,500 K) (Matsui and Abe, 1986) to a warm ocean (500 K) (Fig. 4) yields a secular cooling time of ~$10^3$ years. Latent heat of condensation of water only contributes comparable heat as the sensible heat to the heat budget. Condensation of a steam atmosphere is rapid relative to the timescale of escape (Fig. 5). Following magma ocean crystallization, outgassed $H_2O$ rapidly condenses and the Hadean Earth relaxes into a climate with oceans whose temperature is determined via greenhouse warming by the primordial atmosphere.

## 3.2. Reducing climate model

In this section we introduce a climate model to calculate surface temperatures due to $H_2$ greenhouse warming, which we use in the next section to determine the behavior of hydrogen and deuterium on the Hadean Earth (§4.3). To describe the earliest Hadean climate, we adopt a 2-component ($H_2O$-$H_2$) model for the atmosphere and ocean and take the oxygen fugacity of last equilibration between the magma ocean and primordial atmosphere – and hence the outgassed $H_2/H_2O$ – as a free parameter (Fig. 3). We select a 2-component ($H_2O$-$H_2$) model rather than more complex models (e.g. $H_2O$-CO-$H_2$) to describe a reduced outgassed ocean and atmosphere because: (i) carbon monoxide is not an effective greenhouse gas and is expected to have a secondary influence on calculated surface temperatures (§4.1), (ii) a hydrodynamic hydrogen wind might drag CO to space via molecular collisions and prolong the lifetime of the primordial atmosphere against escape (Zahnle and Kasting, 1986), only strengthening the conclusion that ocean-atmosphere isotopic equilibration was maintained during the escape process (§4.2), (iii) outgassed CO that does not escape is oxidized to $CO_2$ (Kasting, 1990) producing more molecular hydrogen (CO+$H_2O$→$CO_2$+$H_2$). The net effect of incorporating carbon would be to increase the inventory of escaping $H_2$ and deuterium enrichment at any given $fO_2$ of outgassing. However, below we show that even equilibrium deuterium-enrichments calculated by neglecting carbon have sufficient magnitude to yield new constraints on the conditions of primordial outgassing (§4.3). Because water condenses in the lower atmosphere and is retained but molecular hydrogen can escape, the free parameter governing early climate can be expressed as the inventory of molecular hydrogen in the atmosphere ($pH_2$). To translate $H_2/H_2O$ ratios derived from outgassing to atmospheric $H_2$ inventories, we scale by the mass of the ocean reservoir, which we assume to be 1-2 modern ocean equivalents (§3.1)

Due to water condensation and cold-trapping in the lower atmosphere and collision-induced infrared opacity of $H_2$ at moderately high (>0.1 bar) pressures, the emission level in $H_2O$-$H_2$ model atmospheres is determined by the opacity of $H_2$ (Wordsworth, 2012). The emission temperature ($T_E$) is given by top-of-the-atmosphere radiative balance with the early Sun:

$$\frac{L}{4}(1 - A) = \sigma_{SB} T_E^4 \qquad (4)$$

with L the solar constant, A the visible bond albedo, and $\sigma_{SB}$ the Stefan-Boltzmann constant. For L=$10^3$ W/m² appropriate for the early Sun and a bond albedo A=0.3, an emission temperature of 235 K is obtained for the primordial Earth. For simplicity, solar luminosity in these calculations is held constant. Results are qualitatively similar for a range of emission temperatures (215-255 K), as might be expected based on an evolving Sun, cloud feedback and/or by adjusting the planetary albedo to account for Rayleigh scattering in thicker atmospheres. These effects are known to alter the radiation budget by tens of percent (Gough, 1981; Wordsworth, 2012). At infrared wavelengths, the mean optical depth unity surface of a pure $H_2$ atmosphere has been calculated for a several Earth-mass planet (g=20 m/s², $T_{ph}$=100 K) and is ~0.2 bars (Wordsworth, 2012). Combining the expressions for photospheric pressure ($P_{ph} \propto g/\kappa$) and collision-induced opacity ($\kappa \propto \rho$) of an ideal gas ($\rho \propto P_{ph}/T_{ph}$) yields a scaling relation between photospheric pressure, gravity, and photospheric temperature ($P_{ph} \propto g^{1/2} T_{ph}^{1/2}$). Applying this scaling relation to emission temperatures ($T_{ph}$=235K) relevant to the Hadean Earth (g=9.8 m/s²) yields an $H_2$ photospheric pressure ~0.21 bars, which we adopt. Since the atmosphere at the emission level is cold and dry, this pressure – appropriate for a pure $H_2$ atmosphere – is taken as the emission pressure to which a moist adiabatic structure must be stitched. For any given $H_2$ inventory, we use the "all-troposphere" approximation to solve for the surface temperature that yields the emission temperature at the photosphere required by radiative balance (see Supplementary Information). This approximation has been shown to be accurate in describing greenhouse warming in $H_2$-rich atmospheres (Pierrehumbert and Gaidos, 2011).

## 4. Results

In this section, we describe the results of the climate model (§4.1) and use the results to compare the timescales for ocean-atmosphere isotopic equilibration with atmospheric escape (§4.2) to motivate equilibrium isotopic partitioning to describe the behavior of deuterium on the early Earth. Finally, we discuss the hydrogen isotopic evolution of the Hadean oceans and the derived upper limit on $pH_2$ on the early Earth (§4.3).

4.1. Equilibrium surface temperatures

Greenhouse warming by inventories of ~1-$10^2$ bars of equivalent $H_2$ on the Hadean Earth yields surface temperatures ~300-550 K (Fig. 4). Despite some uncertainty arising from cloud feedback or higher albedo due to Rayleigh scattering, multi-bar $H_2$ inventories are sufficient to keep Earth out of a snowball state via $H_2$ opacity alone. The calculation assumes radiative transfer in the terrestrial planet regime, where a sufficient fraction of the visible flux penetrates the atmosphere and powers convection throughout the troposphere, whereas for $pH_2>20$ bars visible photons might not penetrate, creating a deep atmospheric structure governed by geothermal rather than solar heating as in the interiors of the giant planets (Wordsworth, 2012). Calculated surface temperatures for $pH_2>20$ bars are therefore upper limits. These calculations suggest that the dominant control on surface temperatures in the early Hadean is the $H_2$ inventory and that earliest climate on Earth was warm and governed by the physics of atmospheric escape.

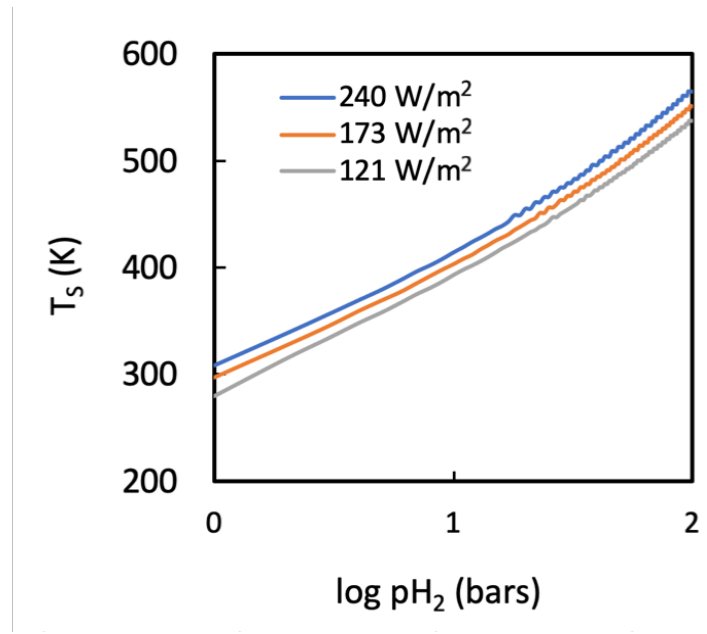

**Figure 4 – Ocean surface temperature as a function of H$_2$ inventory.** Temperatures are shown for different values of the outgoing longwave radiation (OLR) flux for emission temperatures (T$_E$) of 215, 235, and 255 K, corresponding to a range (0-0.5) of bond albedos. The primordial climate depends primarily on the H$_2$ inventory. As has also been found in the case of H$_2$-CO$_2$ early Martian atmospheres (Ramirez et al., 2014), one bar partial pressure of H$_2$ is entirely sufficient to stabilize a warm and wet early climate. The dominant control on earliest climate in these scenarios is the physics of atmospheric escape.

4.2. Atmospheric equilibration and escape timescales

Surface temperatures derived from the climate model can be used to assess H isotopic equilibration between the oceans and atmosphere. A key comparison is between the equilibration time and the residence time of the atmospheric H$_2$ inventory. H$_2$O-H$_2$ deuterium exchange may be rate-limited by exchange in the atmosphere, because H$_2$ dissolves negligibly in the oceans. Deuterium exchange between water vapor and molecular hydrogen is rapid and highly temperature-dependent (Lécluse and Robert, 1994). After condensation of the steam into oceans (§3.1), a significant $H_2$ inventory (>10-100 bars $H_2$) and high surface temperatures (>400-550 K, Fig. 4) result in extremely rapid atmospheric reactions such that the timescale for ocean-atmosphere equilibration is limited by the rate of water evaporation and circulation in a hydrological cycle. This timescale ($\equiv$ depth of oceans/evaporation rate) is rapid ($\approx 10^3$ years), and nominally independent of $H_2$ inventory and surface temperature (e.g. $pH_2$>3 bars, Fig. 5). Ocean-atmosphere isotopic equilibration is therefore expected to initially proceed rapidly. As climate evolves due to $H_2$ escape, reaction rates decline until ocean-atmosphere equilibration becomes rate-limited by exchange reactions in the atmosphere. In this regime, the equilibration time between the oceans and atmosphere is given by:

$$\tau_{AO} = (\tau_{ex}/\tau_{res}) \cdot \tau_{cir} \qquad (5)$$

with $\tau_{ex}$ the isotopic exchange time between atmospheric $H_2$ and $H_2O$, $\tau_{res}$ the residence time of atmospheric water vapor, and $\tau_{cir}$ the timescale to circulate the entire oceans through the atmosphere via evaporation and precipitation in a hydrological cycle (Genda and Ikoma, 2008). As the $H_2$ inventory is lost, the ocean-atmosphere equilibration time ($\tau_{AO}$) evolves for two reasons: (1) the isotopic exchange timescale ($\tau_{ex}$) increases due to the cooling temperatures, (2) the residence time of water vapor in the atmosphere ($\equiv$ depth of equivalent water layer/precipitation rate) decreases due to the lower water vapor pressures associated with lower temperature. Both effects prolong the equilibration time between the ocean and atmosphere, which increases rapidly at low temperatures ($pH_2$<3 bars, Fig. 5), approaching ~1 Myrs for $H_2$ inventories of ~1 bar.

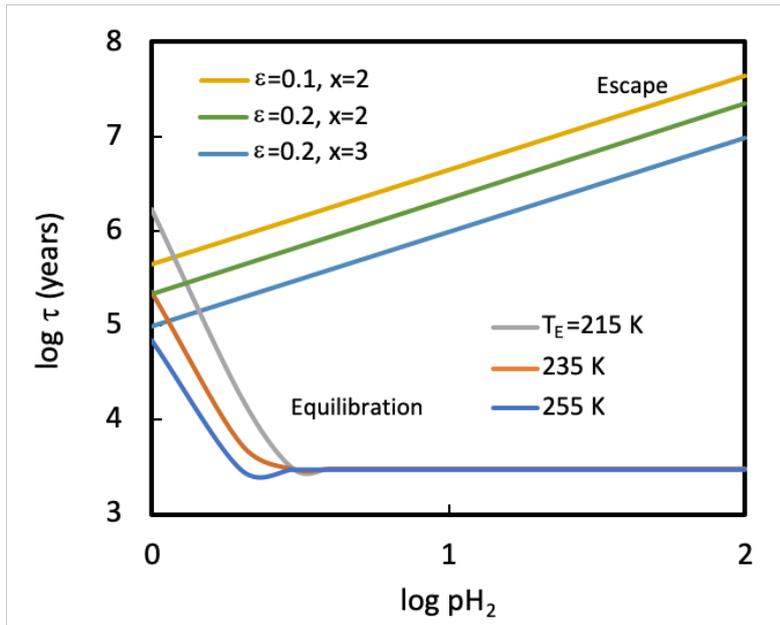

**Figure 5 – Ocean-atmosphere isotopic equilibration and hydrodynamic escape versus H$_2$ inventory.** Isotopic equilibration between the water oceans and a primordial H$_2$-rich atmosphere is calculated for three values of top-of-the-atmosphere emission temperature (T$_E$) as a function of H$_2$ inventory. At high H$_2$ inventories (e.g. 10-100 bars equivalent H$_2$), equilibrium temperatures are high (~400-500K, Fig. 4), and isotopic exchange reactions are extremely fast. Parameters characterizing the escape times are the thermal efficiency ($\varepsilon$) and the planetary EUV absorption radius (x) (§4.2). Timescales for ocean-atmosphere equilibration are generally shorter than the residence time of atmospheric H$_2$ with respect to escape suggesting continuous equilibration during the loss process.

Ocean-atmosphere isotopic equilibration requires that the equilibration time be shorter than the residence time of H$_2$ in the atmosphere. To determine whether or not this is the case, we calculate extreme ultraviolet (EUV) powered escape rates (Watson et al., 1981) for H$_2$ inventories, and assume that loss to space is the sole H$_2$ sink, as expected on the prebiotic Earth. Escape rates can be calculated using the energy-limited approximation:

$$\phi_{H_2} = \frac{\epsilon F_{EUV}}{4R_p} x^2 \qquad (\text{Pa s}^{-1}) \qquad (6)$$

with $\phi$ the hydrogen escape flux expressed as atmospheric pressure loss per unit time (Wordsworth, 2012), F$_{EUV}$ the extreme ultraviolet flux of the young sun, ε the thermal efficiency or the fraction of incident EUV used to power the planetary wind (ε=0.1-0.2), R$_p$ the planetary radius, and x($\equiv$R$_{EUV}$/R$_p$) the effective absorption radius in planetary radii (x=2-3), which characterizes the distended nature of EUV absorption in escaping atmospheres. F$_{EUV}$ is assumed equal to 100 times the modern extreme ultraviolet flux, i.e. 464 erg cm$^{-2}$ s$^{-1}$ (Ribas et al., 2005). Other parameter choices (ε,x) are adopted from simulations of hydrogen-rich atmospheres exposed to comparable EUV fluxes (Erkaev et al., 2016; Shematovich et al., 2014). Calculated residence times ($\equiv$pH$_2$/$\phi_{H2}$) for H$_2$ are ≈ 10$^6$-10$^7$ years for H$_2$ inventories ≈ 10-100 bars (Fig. 5). Timescales for ocean-atmosphere equilibration are generally shorter than the residence time of atmospheric H$_2$, suggesting continuous equilibration with ocean-atmosphere quenching occurring at H$_2$ inventories of a few bars or less.

4.3. Hydrogen isotopic evolution of the Hadean oceans

Finally, we quantify the behavior of hydrogen ($^1$H) and deuterium ($^2$D) as tracers of early Earth evolution to articulate constraints on the chemical composition of the primordial atmosphere. In brief, water vapor is retained via condensation but hydrogen in non-condensable form (e.g. H$_2$) interacts with water vapor in the lower atmosphere and is transported to the upper atmosphere and lost to space. Equilibrium D/H partitioning between water and hydrogen is calculated from standard prescriptions (Richet et al.,

1977).[1] Although this prescription strictly relates water vapor to molecular hydrogen, it can be used to characterize equilibrium between water oceans and $H_2$-rich atmospheres because the vapor pressure isotope effect relating liquid water to water vapor is an order of magnitude smaller and can be neglected to first order. $H_2O$-$H_2$ equilibration is among the largest equilibrium fractionations between two molecules in nature, with a clearly resolvable magnitude at planetary temperatures. D/H composition of planetary oceans reflect the mass of early $H_2$ reservoirs and the temperatures of isotopic equilibration. To the extent that an atmospheric $H_2$ reservoir comparable to the hydrogen in the terrestrial oceans was present, the ocean D/H could have evolved dramatically due to equilibrium partitioning and removal of isotopically light $H_2$.

The hydrogen isotopic evolution of Hadean oceans can be calculated using equilibrium ocean-atmosphere partitioning. Following magma ocean crystallization, most of the water vapor condenses into oceans, while most $H_2$ partitions into the atmosphere, generating the earliest climate (§3). The magnitude of greenhouse warming for a freshly outgassed atmosphere is significant: only a few percent of the outgassed inventory need be in the form of molecular hydrogen to prevent a snowball Earth and to stabilize water oceans via $H_2$ greenhouse warming alone (Fig. 4). The existence of such an early greenhouse climate permits isotopic equilibration between the ocean and atmosphere (§4.2) with the temperature-dependent partitioning between reservoirs determined self-consistently via climate with a given $H_2$ inventory (§3). As the atmospheric $H_2$ inventory is depleted via escape, the greenhouse effect also diminishes, accentuating the temperature-dependent partitioning between ocean and atmosphere. Hence, deuterium is further concentrated into the oceans due to the cooling radiative balance accompanying

---

[1] The exchange reaction is $H_2O + HD \Leftrightarrow HDO + H_2$ with equilibrium constant $K(T) = 1 + 0.22*(10^3/T)^2$.

$H_2$ loss. In this way, D/H evolution of the oceans can be calculated as an equilibrium distillation sequence, converging to a value determined by the initial inventory of $H_2$. Water molecules have a strong tendency to concentrate deuterium, and, to the extent that $H_2$ is abundant in the primordial atmosphere, pronounced D/H enrichments (~1.5-2x) can arise from equilibration and $H_2$ escape in this epoch (Fig. 6), an enrichment not evident in Earth's chondritic oceans. Such deuterium enrichment is minimized for a pure steam atmosphere outgassed from an oxidizing magma ocean. Preservation of a chondritic D/H signature in the oceans to within <20% thereby constrains $H_2/H_2O$ of Earth's outgassed atmosphere to <0.3. For a given $H_2O$ abundance, e.g., 1-2 modern ocean equivalents (Korenaga, 2008), upper limits on the initial $H_2$ inventory of Earth ($pH_2$<10-20 bars) can be derived (Fig. 6).

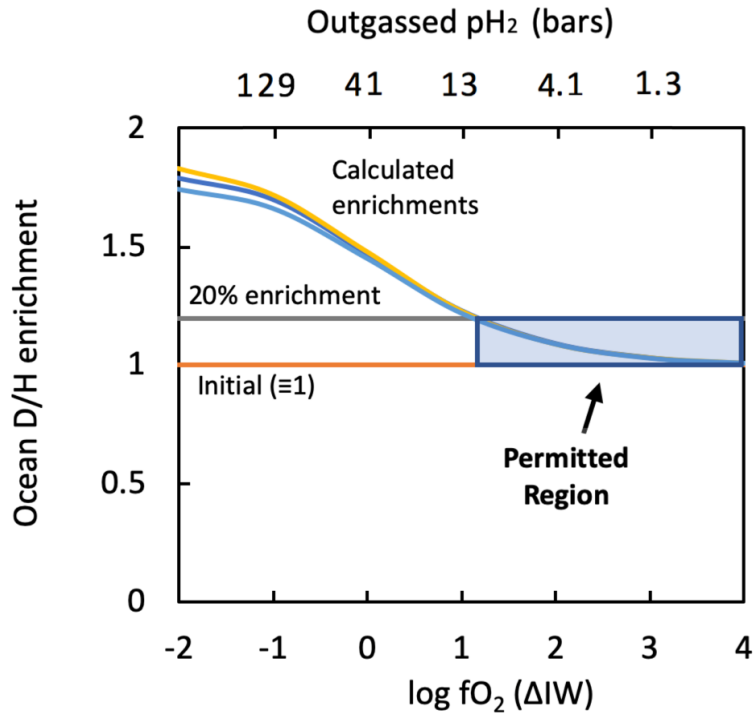

**Figure 6 – Ocean deuterium enrichment versus oxygen fugacity of primordial outgassing.** Oxygen fugacity determines the $H_2/H_2O$ of the outgassed atmosphere. Oxidizing conditions lead to nearly pure steam atmospheres and minimal ocean D/H enrichment whereas reducing outgassing (e.g., $logfO_2<IW$) generates higher $H_2/H_2O$ (>1) and stronger ocean D/H enrichments (>1.5-2x) due to equilibration and loss of large quantities of deuterium-depleted $H_2$. Minimal (<20%) deuterium-enrichment in terrestrial water relative to the source (§5) constrains $H_2/H_2O$ of the outgassed atmosphere to <0.3 and $logfO_2$ of outgassing to >IW+1 (Permitted region). Tolerable $H_2$ abundances can be expressed in absolute terms (top axis, $pH_2$<10-20 bars) by fixing water abundances to 1-2 modern ocean equivalents. Temperatures of ocean-atmosphere equilibration are calculated via the climate model (§3.2), with the enrichment curves corresponding to different emission temperatures ($T_E$=215, 235, 255K), demonstrating the robustness of the result to plausible variations in early climate.

Here, we have only considered ocean-atmosphere partitioning followed by the removal of atmospheric $H_2$ with no isotopic fractionation during the escape process. Kinetic processes (e.g. $HD/H_2$ mass-fractionation) could fractionate H isotopes further (Zahnle et al., 1990) and cause additional ocean deuterium enrichment for a given $H_2$ inventory although a sufficiently vigorous hydrogen escape flux would produce a subdominant or negligible kinetic contribution to the D/H enrichment relative to equilibrium partitioning (Genda and Ikoma, 2008). The constraints on primordial $H_2$ abundances shown in Figure 6 are therefore strictly speaking upper limits. We compare these results to those obtained previously. For initial outgassed envelopes with $H_2/H_2O$=1-10, ocean D/H enrichments of 2-3x due to equilibrium partitioning alone and 2-9x by inclusion of $HD/H_2$ kinetic mass fractionation during escape have been reported (Genda and Ikoma, 2008). These previous results were obtained by assuming ocean-atmosphere equilibration occurred at 300 K for all $H_2$ inventories ($pH_2 \simeq$30-300 bars for initial $H_2/H_2O$=1-10). By contrast, we find higher temperatures of ocean-atmosphere equilibration ($T_s$=475-550 K for $pH_2$=30-100 bars, Fig. 4). By calculating climate and temperature-dependent isotopic partitioning self-consistently, we expect our model to yield a more accurate description of the equilibrium behavior of deuterium on the Hadean Earth. We find ocean deuterium enrichments of 1.5-1.8x (Fig. 6) for initial $H_2$ inventories of 30-300 bars ($H_2/H_2O \approx$1-10) corresponding to reducing conditions for primordial outgassing ($\log fO_2$=IW to IW-2). We do not attempt to estimate kinetic mass fractionation during escape but note that the near constancy of terrestrial D/H in the last 3.8 Ga (Pope et al., 2012) restricts any primordial $H_2$ inventory to the Hadean Earth and implies a vigorous hydrogen wind with a subdominant role for kinetics (Genda and Ikoma, 2008). Calculated enrichments are therefore a conservative but robust constraint on minimum ocean deuterium enrichment for a given past H$_2$ inventory. In summary, calculated ocean D/H enrichments (1.5-1.8x, Fig. 6) are somewhat smaller than those predicted by (Genda and Ikoma, 2008) via equilibrium partitioning alone (2-3x, their Fig. 6) because greenhouse warming reduces equilibrium isotopic fractionation and permits more deuterium to partition into escaping H$_2$ in our calculations. Neither our calculated equilibrium enrichments (1.5-1.8x) nor those reported previously (2-3x) (Genda and Ikoma, 2008) for H$_2$-rich outgassing are sufficiently high to reconcile a nebular source with the terrestrial oceans, supporting the carbonaceous chondrite origin of the terrestrial hydrosphere. Nevertheless, calculated ocean deuterium enrichments are sufficiently large to yield new constraints on Hadean evolution, which we discuss next.

## 5. Discussion

The molecular composition of Earth's primordial atmosphere is not well-constrained. Nevertheless, on the basis of their isotopic compositions, Earth's major volatiles (H, N, C) are thought to be sourced primarily from carbonaceous chondrites (Alexander et al., 2012; Halliday, 2013; Marty, 2012). This widely-held view of the source of major terrestrial volatiles requires preservation of the source signature in the terrestrial oceans and implies minimal D/H enrichment via equilibration and escape of primordial H$_2$. To quantify the constraint that this comparison places on primordial outgassing, we compare the isotopic composition ($\delta$D=-25 +/- 5‰) of the Archean oceans (Pope et al., 2012) with the lowest bulk chondritic values ($\delta$D=-226 +/- 4‰) measured to date (Alexander et al., 2012). On this basis, a primarily chondritic source for terrestrial water requires <20% deuterium-enrichment via H$_2$ loss. According to our calculations, this level of isotopic preservation requires most early outgassed hydrogen from the Earth to appear in the form of water vapor (H$_2$/H$_2$O<0.3, pH$_2$<10-20 bars, Fig. 6). By connecting the conditions of outgassing to the observable isotopic signatures in ancient seawater, we can articulate new constraints on the composition of the Earth's primordial atmosphere. These results show that reducing gases such as $H_2$ and $CH_4$ made up only a minor fraction of the Earth's outgassed atmosphere and require the terminal magma ocean to be oxidized by the time of last equilibration with the atmosphere. In this section, we discuss the implications of these results for redox evolution of the magma ocean (§5.1) and the oxidant involved in terrestrial late accretion (§5.2).

5.1. The redox state of the terrestrial magma ocean

The redox state of a magma ocean determines both the chemical composition of the outgassed atmosphere and the isotopic composition of water oceans following primordial $H_2$ loss. Oxidizing magma oceans outgas water-rich primordial atmospheres, which condense into oceans, experiencing minimal hydrogen escape and deuterium enrichment. Reducing magma oceans, by contrast, outgas substantial quantities of hydrogen as $H_2$ in addition to water molecules whose equilibration with water oceans before loss can enrich the oceans in deuterium by ~1.5-2x relative to initial values, a feature not evident in the terrestrial isotopic record (Pope et al., 2012). The persistence of a chondritic signature in the terrestrial oceans requires a low outgassed $H_2/H_2O$ (<0.3) and oxidizing conditions for last atmospheric equilibration with the magma ocean ($\log fO_2 > IW+1$) (Fig. 6). Given that the convective magma ocean initially held metallic droplets in suspension (Stevenson, 1990) and was therefore chemically reducing ($\log fO_2 < IW-2$) at early times, these results suggest that the silicate Earth was oxidized *during* the evolution of the magma ocean. Three mechanisms have been discussed for this primordial oxidation: (1) the terrestrial magma ocean could have been oxidized via Fe disproportionation at high pressure ($3Fe^{+2} \rightarrow 2Fe^{+3} + Fe^0$) with separation of the newly generated metallic iron to the core leaving an oxidized mantle residue (Hirschmann, 2012; Wade and Wood, 2005), (2) the primordial atmosphere was reducing ($H_2$-rich) but the process of $H_2$ escape during the lifetime of the magma ocean oxidized both the atmosphere and the co-existing silicate Earth (Hamano et al., 2013), and (3) the $Fe^{+3}/Fe^{+2}$ value of the terrestrial magma ocean was low (~0.01) but the more incompatible nature of ferric iron ($Fe^{+3}$) in mantle minerals enriched it in evolving liquids such that the late-stage magma ocean was more oxidizing than that at the outset of crystallization (Schaefer et al. submitted). The relative importance of these processes for the redox evolution of magma oceans is subject to future study. For now, we conclude that the oxidation of the silicate Earth occurred during the crystallization of the magma ocean, independently corroborating the conclusion from geological data for early (>3.5-4 Gya) establishment of oxidizing conditions in Earth's mantle (Delano, 2001; Nicklas et al., 2018; Trail et al., 2011).

5.2. The oxidant for terrestrial late accretion

Before outgassing of the primordial atmosphere, the magma ocean potentially facilitates the last major episode of core formation via separation of metallic droplets accompanying deep magma ocean convection on rapid (~$10^2$ year) timescales (Stevenson, 1990). Such metallic droplets strongly concentrate and efficiently scavenge highly siderophile elements (HSEs) from the terrestrial magma ocean and sequester them into the metallic core. Mantle relative abundances of HSEs resemble the chondrites, leading to the notion that these elements were delivered during the final ~1% of Earth accretion, after cessation of core formation (Kimura et al., 1974), now interpreted as accretion after the Moon-forming giant impact. Isotopic characteristics of the Earth's mantle suggest delivery by bodies with metals either as undifferentiated metallic grains or as planetesimal cores (Marchi et al., 2018). However, the terrestrial upper mantle is currently unsaturated in metallic Fe, instead exhibiting a more oxidizing redox state, indicated by higher $Fe^{+3}/Fe^{+2}$ values than those characterizing co-existence with metallic iron. Accordingly, accreted metals must have been oxidized and dissolved into Earth's mantle, prompting the question of the nature of the oxidant involved in late accretion. Since the terrestrial magma ocean crystallized on $\sim10^6$ year timescales (Lebrun et al., 2013) while the leftovers of planetary accretion were swept up over $\sim10^7$-$10^8$ years (Morbidelli et al., 2012), late accretion likely occurred onto a silicate Earth with water oceans. Possibilities for oxidizing the metals delivered during late accretion are: (1) Earth's fluid envelope, e.g., water in the terrestrial oceans, via the iron-water reaction ($Fe+H_2O \rightarrow FeO+H_2$) followed by hydrodynamic escape of $H_2$ (Genda et al., 2017) and (2) the oxidizing power of Earth's mantle, epitomized by ferric iron ($2Fe^{+3}+Fe^0 \rightarrow 3Fe^{+2}$), lowering the ferric iron abundance to the modern upper mantle value ($Fe^{+3}/Fe^{+2}$=0.03-0.04) (Canil et al., 1994). Using calculated enrichments in the D/H of water oceans coexisting with significant ($\sim$10-100 bar) early $H_2$ atmospheric inventories, we can limit the extent of the iron-water reaction in oxidizing the metals of late accretion. Given that the oxidation of metals at that time would consume more than a modern ocean worth of water via the iron-water reaction (producing >30 bars $H_2$) and that the Hadean $H_2$ inventory was apparently <10-20 bars (Fig. 6), we conclude that the iron-water reaction had a subdominant role during late accretion. A more dominant role for this reaction would have produced deuterium-enriched oceans not observed in the terrestrial record. This reasoning suggests that the terrestrial mantle supplied oxidants during late accretion, a feature that may yield insights into the physics and chemistry of this early process.

## 6. Conclusions

The isotopic composition of the oceans provides a unique constraint for early planetary evolution. It is widely accepted that most water accreted by the Earth was delivered before the Moon-forming giant impact and that most water dissolved in the subsequent magma ocean was excluded from crystallizing minerals and outgassed into a primordial atmosphere (Elkins-Tanton, 2008; Greenwood et al., 2018). Given that the residence time of water in Earth's oceans with respect to the deep water cycle is comparable to, or greater than, the current age of the Earth (van Keken et al., 2011), most of the hydrogen atoms in the oceans today are inferred to be outgassed from the magma ocean, retaining isotopic memory of early epochs.

The oxygen fugacity of terrestrial magma ocean outgassing – and therefore the chemical composition of the primordial atmosphere – has not been independently constrained. By linking the oxygen fugacity of primordial outgassing to the deuterium content of Earth's hydrosphere, we articulate new constraints on these critical parameters governing early Earth evolution. We find that preservation of a carbonaceous chondritic D/H signature in the terrestrial oceans (to 10-20%) requires Earth's outgassed envelope be hydrogen-poor ($H_2/H_2O<0.3$), indicating oxidizing conditions ($logfO_2>IW+1$) for last equilibration with the magma ocean. We infer that oxidation of the silicate Earth took place during the evolution of Earth's final magma ocean, and may require no geological oxidation processes (e.g. subduction) to be consistent with an oxidized mantle observed in the earliest terrestrial record (Trail et al., 2011).

The inferred absence of massive (>20 bar) $H_2$ inventories of any origin on the Hadean Earth constrains the oxidant for terrestrial late accretion. Whereas the likely existence of early water oceans has previously been taken to imply that the iron-water reaction was responsible for oxidizing the metals delivering HSEs to early Earth (Genda et al., 2017), we find that such massive production of molecular hydrogen would have disturbed the carbonaceous-chondrite-like signature of the terrestrial oceans. We therefore infer that oxidants in the terrestrial mantle (e.g., $Fe^{+3}$) were responsible for oxidative resorption of late-accreting metals delivered to the silicate Earth. Indeed, the oxidative potential in Earth's modern mantle is comparable to the reducing potential in ~0.5% of an Earth mass of chondritic late accretion, a feature that may yield insight into this early terrestrial process.

**Acknowledgements**

K.P. acknowledges support from a grant from the W.M. Keck foundation. The authors are grateful to Peter Buseck for detailed comments on an early draft, and to Fabrice Gaillard and two anonymous reviewers for thorough reviews that greatly helped to improve the manuscript.

**Supplementary Information – Calculation of surface temperatures and moist adiabats**

Planetary surface temperature is controlled by the structure of the troposphere. At the base of the troposphere, vapor pressure equilibrium with the ocean controls the water vapor abundance. We assume a troposphere saturated in water vapor throughout. Accordingly, the partial pressure of water vapor is given by $p_{H2O} = \exp(-\Delta G/RT)$, with $\Delta G=\Delta H-T\Delta S$ and $\Delta H$=40.58 kJ/mol, $\Delta S$=0.1082 kJ/mol.K (Chase et al., 1985). As in any multi-component atmosphere, the inventory of one gas influences the partial pressure of other gases through vertical redistribution (Wordsworth and Pierrehumbert, 2013). The total pressure is assumed given by the expression $p_T = \sigma_{H2O} g\mu/\mu_{H2O} x_{H2O}$ with $\sigma_{H2O}$ the surface density of water vapor, g the gravitational acceleration at the surface, µ the mean molecular weight, and $\mu_{H2O}$ and $x_{H2O}$ the atomic mass and the mole fraction of water vapor. This expression is strictly valid in a well-mixed atmosphere, which we take as an adequate approximation for an $H_2O$-$H_2$ atmosphere. With the above relations, we write an expression for the entropy of atmospheric gas in contact with the oceans, which we use to calculate surface temperature. The entropy of the troposphere is that of the basal gaseous mixture:

$$S(T) = \sum_i x_i s_i(T) - R\sum_i x_i \ln x_i - R\ln P \qquad (A1)$$

which is the expression for entropy of an ideal mixture of ideal gases, with the first term the sum over species as pure gases at standard pressure, the second term an entropy of mixing term deriving from the fact that the gas parcel is a mixture of randomly distributed gas molecules, and the third term a pressure correction due to the volume available to each molecule. In this way, the entropy of the atmospheric parcel at any temperature (T), pressure (P) and composition ($x_i$) can be calculated. We consider ideal gas theory an adequate approximation for the primordial atmosphere because intermolecular distances are large relative to the size of the molecules. The thermodynamic data for the entropy of pure substances is taken from standard sources (Chase et al., 1985).

Because a moist adiabat is also isentropic, the tropospheric entropy can be used to relate the conditions at the base to those characterizing the radiative emission level where the mode of energy transport transitions to radiation. A major influence on tropospheric structure is the condensation of water vapor into clouds through adiabatic expansion and cooling. The specific entropy at the radiative emission level at the top of the troposphere must therefore consider condensates:

$$S(T) = F_v S_v(T) + F_L S_L(T) \tag{A2}$$

with $F_V$ and $F_L$ the fraction of total molecules in the parcel in the gas and condensates, respectively. Atmospheric P-T paths determined by calculating pseudoadiabats (i.e., with rainout) are similar to adiabatic equivalents (i.e., with condensates suspended) (Ingersoll, 1969), permitting use of a two-phase isentrope even when rainout might be expected and upper tropospheric opacity determined by gas-phase ($H_2$) opacity alone. The procedure for calculating surface temperature entails: (1) an initial estimate for surface temperature, yielding the partial pressure of water vapor and, with a given $H_2$ inventory, total surface pressure, (2) calculation of the entropy of the convective atmosphere, (3) evaluation of the thermodynamic state, including temperature, of the atmospheric parcel at a pressure of 0.21 bars representing the radiative emission level. In this way, each value of surface temperature corresponds to an emission temperature, and iteration allows identification of the surface temperature corresponding to the emission temperature required by top-of-the-atmosphere radiative balance (Equation 4). Because the convective troposphere is isentropic, we can solve for the surface temperature as a function of emission temperature without explicitly resolving the vertical structure in the intervening atmosphere. In this way, we calculate surface temperatures in the computationally efficient "all-troposphere" approximation (Pierrehumbert, 2010) and iterate to find solutions.